\documentclass[twoside,twocolumn,9pt]{article}
\usepackage{extsizes}
\usepackage[super,sort&compress,comma]{natbib} 
\usepackage[version=3]{mhchem}
\usepackage[left=1.5cm, right=1.5cm, top=1.785cm, bottom=2.0cm]{geometry}
\usepackage{balance}
\usepackage{mathptmx}
\usepackage{sectsty}
\usepackage{graphicx} 
\usepackage{lastpage}
\usepackage[format=plain,justification=justified,singlelinecheck=false,font={stretch=1.125,small,sf},labelfont=bf,labelsep=space]{caption}
\usepackage{float}
\usepackage{fancyhdr}
\usepackage{fnpos}
\usepackage[english]{babel}
\addto{\captionsenglish}{%
}
\usepackage{array}
\usepackage{droidsans}
\usepackage{charter}
\usepackage[T1]{fontenc}
\usepackage[usenames,dvipsnames]{xcolor}
\usepackage{setspace}
\usepackage[compact]{titlesec}
\usepackage{hyperref}

\usepackage{epstopdf}

\definecolor{cream}{RGB}{222,217,201}

\begin{document}

\pagestyle{fancy}
\thispagestyle{plain}
\fancypagestyle{plain}{
\renewcommand{\headrulewidth}{0pt}
}

\makeFNbottom
\makeatletter
\renewcommand\LARGE{\@setfontsize\LARGE{15pt}{17}}
\renewcommand\Large{\@setfontsize\Large{12pt}{14}}
\renewcommand\large{\@setfontsize\large{10pt}{12}}
\renewcommand\footnotesize{\@setfontsize\footnotesize{7pt}{10}}
\makeatother

\renewcommand{\thefootnote}{\fnsymbol{footnote}}
\renewcommand\footnoterule{\vspace*{1pt}%
\color{cream}\hrule width 3.5in height 0.4pt \color{black}\vspace*{5pt}} 
\setcounter{secnumdepth}{5}

\makeatletter 
\renewcommand\@biblabel[1]{#1}            
\renewcommand\@makefntext[1]%
{\noindent\makebox[0pt][r]{\@thefnmark\,}#1}
\makeatother 
\renewcommand{\figurename}{\small{Fig.}~}
\sectionfont{\sffamily\Large}
\subsectionfont{\normalsize}
\subsubsectionfont{\bf}
\setstretch{1.125} 
\setlength{\skip\footins}{0.8cm}
\setlength{\footnotesep}{0.25cm}
\setlength{\jot}{10pt}
\titlespacing*{\section}{0pt}{4pt}{4pt}
\titlespacing*{\subsection}{0pt}{15pt}{1pt}


\fancyfoot{}
\fancyfoot[C]{\footnotesize\sffamily\thepage}

\fancyhead{}
\renewcommand{\headrulewidth}{0pt}
\renewcommand{\footrulewidth}{0pt}

\makeatletter 
\newlength{\figrulesep} 
\setlength{\figrulesep}{0.5\textfloatsep} 

\newcommand{\topfigrule}{\vspace*{-1pt}%
\noindent{\color{cream}\rule[-\figrulesep]{\columnwidth}{1.5pt}} }

\newcommand{\botfigrule}{\vspace*{-2pt}%
\noindent{\color{cream}\rule[\figrulesep]{\columnwidth}{1.5pt}} }

\newcommand{\dblfigrule}{\vspace*{-1pt}%
\noindent{\color{cream}\rule[-\figrulesep]{\textwidth}{1.5pt}} }

\makeatother

\twocolumn[
  \begin{@twocolumnfalse}
{
}\par
\vspace{1em}
\sffamily

 \noindent\LARGE{\textbf{Field-Directed Assembly of Anisotropic Composite `Pacman' Colloids into Reconfigurable Architectures}} \\

  \noindent\large{Indira Barros\textit{$^{a}$}, Isha Songara\textit{$^{a}$} and Indrani Chakraborty$^{\ast}$\textit{$^{a}$}} \\

 \noindent\normalsize{We investigate the field-directed assembly of patchy colloidal particles composed of magnetic hematite cubes encapsulated in TPM shells. These “Pacman-like” particles exhibit specific interactions via their magnetic patch, enabling tunable structures under external fields. A magnetic field drives patch-to-patch interactions, while an AC electric field induces stacking and spreading. By using this multi-field combination and by varying field parameters, such as frequency and amplitude we achieve programmable transitions between distinct assembly modes. Our work offers a versatile platform for designing reconfigurable colloidal architectures with potential applications in micro-robotics and smart materials.} \\


 \end{@twocolumnfalse} \vspace{0.6cm}

  ]

\renewcommand*\rmdefault{bch}\normalfont\upshape
\rmfamily
\section*{}
\vspace{-1cm}


\footnotetext{\textit{$^{a}$~Department of Physics, Birla Institute of Technology and Science, Pilani, K K, Birla Goa Campus, Zuarinagar, Sancoale, Goa 403726, India. \\
*E-mail: indranic@goa.bits-pilani.ac.in}}





\section{Introduction}

Multi-functional patchy, Janus and composite particles have become an area of intense research in the last decade. These versatile particles offer enhanced tunability and functionality within the same system, thereby providing us with building blocks for the programmable assembly of complex, reconfigurable, and stimuli-responsive structures, as well as fabrication of micro-swimmers for targeted drug delivery. \cite{chen2025micromotors, hou2023review} Integration of these composite particles with multi-field control, for example, assembly and/or propulsion using electric, magnetic and optical fields, has further accelerated the promise of real-world applications, for example, propulsion inside biological systems, sensing of target biomolecules and tunable mechanical metamaterials.\cite{chuang2025periodic, liu2024programmable, campuzano2019magnetic}

An immensely interesting composite system in this context is the hematite ($\alpha$-Fe$_2$O$_3$) cube embedded in a spherical shell of 3-(trimethoxysilyl)propyl methacrylate (TPM).\cite{sacanna2012magnetic} When suspended in water, without any applied fields, these TPM-hematite composites interact primarily through their magnetic cube region that possesses a residual magnetic moment, while the rest of the surface remains inert. Like the arcade character Pacman, only the ``mouth''- here, the magnetic patch- is active, leading to assemblies that resemble Pacman-like units clustering via their functional sites. Therefore, we refer to this composite system as the `Pacman particle'. This composite system offers structural asymmetry, localized magnetic interactions, and photocatalytic activity emerging from the cubic hematite patch. The localized magnetic interactions can be exploited to direct colloidal assembly, as demonstrated in the work of Sacanna and co-workers on magnetic click colloidal assembly, where magnetic interactions between colloidal building blocks enable the formation of organized structures.\cite{sacanna2012magnetic} As hematite is a visible-bandgap semiconductor, illuminating an exposed hematite face with blue light in the presence of H$_2$O$_2$ drives a photocatalytic breakdown of H$_2$O$_2$ into water and oxygen. Since this happens only on the hematite side of the composite particle, a self-phoretic motion is induced that underlies the living, light-switchable colloidal crystals of Palacci \textit{et al.} \cite{palacci2013living}, the phoretic ``docker'' particles that capture, tow, and release inert TPM cargo under combined light and magnetic steering, \cite{palacci2013photoactivated}, and subsequent multiwavelength extensions of the same activation strategy \cite{palacci2014light, palacci2015artificial}.

The area that remains entirely unexplored is the response of the Pacman particle to an external (AC) electric field, or a combination of electric and magnetic fields. In contrast to optically activated systems, magnetic and electric-field-driven approaches offer significantly higher potential for practical, real-world applications. This is because magnetic field-driven systems can be operated inside biological environments due to the easy penetrability of magnetic fields inside biological tissues, and electric field-driven microswimmers do not require locally available photocatalytic fuels such as H$_2$O$_2$. The Pacman particle offers multi-field controllability within a single system, enabling independent or combined manipulation using both magnetic and electric fields. Hematite ($\alpha$-Fe$_2$O$_3$) is a canted antiferromagnet, with a small permanent magnetic dipole moment at room temperature.\cite{morrish1994canted} TPM is a non-magnetic material, so a conjugate of a TPM sphere and a hematite cube produces a non-magnetic particle with a magnetic `patch'. On the other hand, the hematite–TPM conjugate is dielectric, and an applied AC electric field induces a dipole moment whose magnitude and direction are governed by the frequency-dependent Clausius–Mossotti factor, which reflects the contrast in the complex permittivities of the particle and the surrounding medium. For simple dielectric spheres under applied AC electric fields, a complex interplay of dipolar, dielectrophoretic and electrohydrodynamic forces gives rise to a diverse range of structures from chains, two- and three-dimensional colloidal crystals to colloidal oligomers, to hierarchical colloidal chains and clusters. \cite{vutukuri2012colloidal, leunissen2009directing, ma2013formation, ma2015electric, gong2017electric, lumsdon2004two, trau1996field, juarez2011electric, barros2025structure} Unlike a homogeneous dielectric sphere, the hematite-cube-in-TPM-sphere particle is a genuinely composite dielectric object: hematite is a semiconducting oxide whose permittivity is large and strongly frequency- and size-dependent \cite{lunt2013dielectric, papaioannou2005electron}, while the surrounding TPM shell is an organosilica polymer with a permittivity much closer to that of conventional silicate glasses. An AC electric field applied to such a particle should therefore not induce a single, spatially uniform dipole, as it would for a homogeneous sphere, but rather a spatially structured polarization response. Therefore, a systematic investigation of the response of Pacman particles to multi-field excitations is both necessary and important for unlocking their full potential for programmable assembly.

In this work, we perform a systematic investigation of the self-assembly and collective dynamics of the Pacman particles under single-field and multi-field excitations. By varying the relative orientation of the applied electric and magnetic fields between parallel and orthogonal configurations, we investigate how the competition between electric and magnetic interactions drives the system toward distinct final structural configurations. Our system simultaneously offers three things: a) global reconfigurability by electric field-induced assembly, b) local reorientation by applied magnetic field, and c) structural anisotropy for generating flow asymmetry. This makes Pacman particles an immensely promising candidate as building blocks for reconfigurable smart materials and for building field-controlled microswimmers. 

\section{Materials and Methods}

\begin{figure}[h]
	\centering
	\includegraphics[width=8.3 cm]{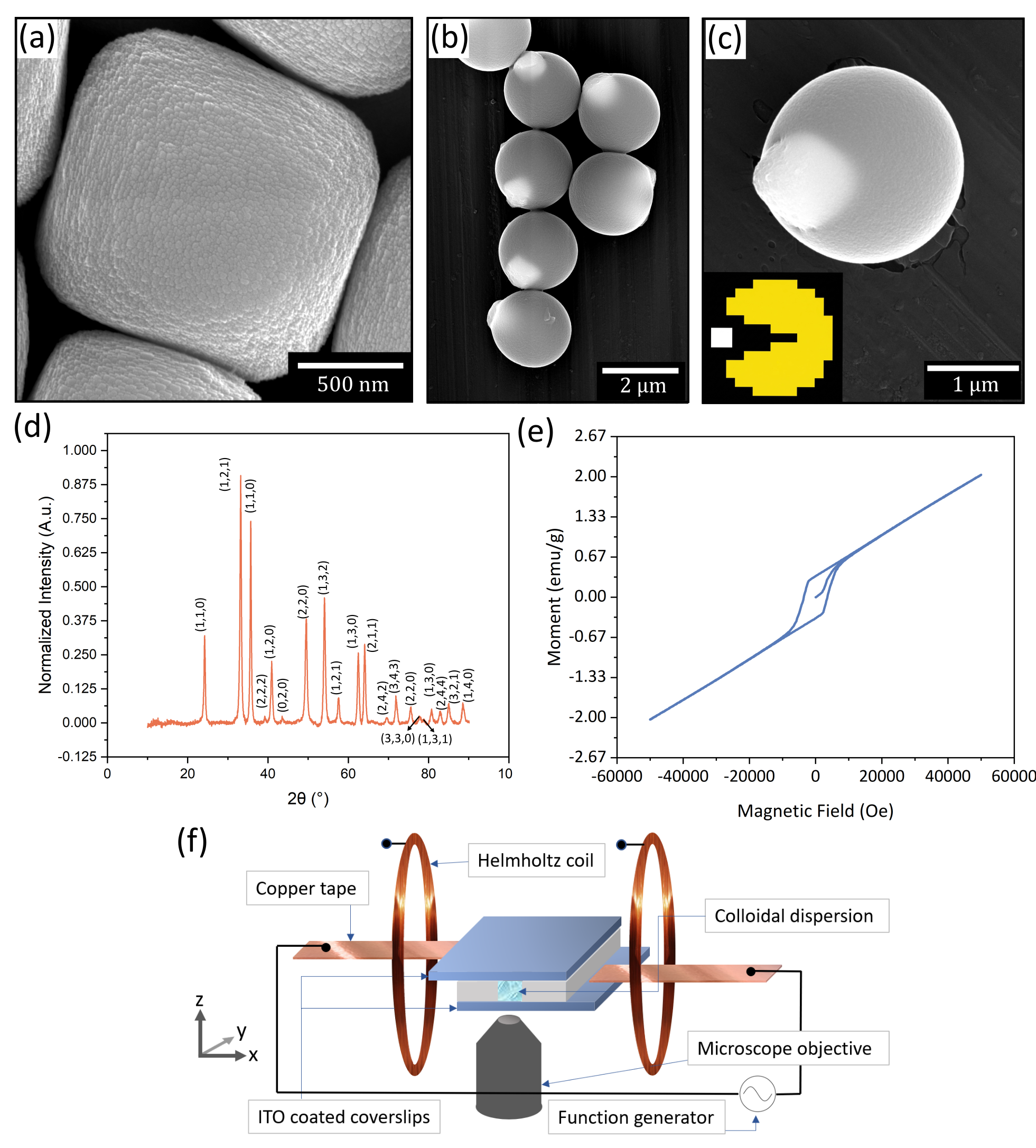}
	\caption{(a) SEM images of hematite cubes synthesised by a sol-gel method. (b) The hematite cubes encapsulated in TPM spheres, making the cube + TPM conjugate: a ‘Pacman’ particle. (c) A Pacman particle with an inset of a schematic of the arcade character, Pacman. (d) X-ray diffraction pattern of the hematite cubes with each peak marked out by the respective (hkl) values, which matches the trigonal phase of $\alpha-Fe_2O_3$. (e) The magnetization vs applied magnetic field (M-H) plot of the hematite cubes shows well-defined hysteresis, indicating the canted antiferromagnetic nature of hematite at room temperature. (f) Schematic of the experimental setup and sample geometry used for the field-based experiments.}
	\label{fig1}
\end{figure}

\subsection{Particle synthesis}
The synthesis, developed by Sacanna \textit{et al.}, proceeds by heterogeneous nucleation and controlled wetting of hydrolyzed TPM oil on a pre-formed hematite seed, so that the extent to which the oil engulfs the seed, and hence the final particle geometry, from a bare cube to a fully embedded off-centre core, can be tuned through the pH and the relative amounts of hematite and TPM \cite{sacanna2012magnetic, van2017preparation, youssef2016shape, palacci2015artificial}. Since the hematite core itself can be synthesized over a wide range of shapes and sizes \cite{sugimoto1992preparation}, this single protocol has produced a whole family of building blocks,\cite{meijer2021preparation} and has become a standard platform for studies of cubic superball crystallization \cite{rossi2015shape}, dipolar cube fluids \cite{rossi2018self}, two-dimensional chiral fluids \cite{soni2019odd}, microroller fronts and swarms \cite{driscoll2017unstable}, and self-spinning, diffusiophoretically coupled microgears \cite{aubret2018targeted, aubret2018diffusiophoretic}. 

Hematite cubes of size $1.0 \pm 0.1\,\mu\mathrm{m}$ were synthesised by this sol-gel method (Fig.~\ref{fig1}a), as outlined by Sugimoto \textit{et al}\cite{sugimoto1992preparation}. X-ray diffraction studies of the cubes showed a correspondence to the trigonal phase of $\alpha$-Fe$_2$O$_3$ (Fig. \ref{fig1}d). The particles exhibit canted antiferromagnetism,\cite{morrish1994canted} so there is a residual magnetic moment as shown in the M-H plot (Fig. \ref{fig1}e). For encapsulation of hematite cubes by TPM shells, $60\,\mu\mathrm{L}$ of aqueous ammonia (28~wt\%) was added to $60\,\mathrm{mL}$ of an aqueous suspension of cubes (0.5~wt\%). After thorough mixing, $400\,\mu\mathrm{L}$ of the oil precursor, 3-(methacryloxypropyl)trimethoxysilane (TPM), was introduced into the suspension under continuous stirring. The nucleation and subsequent growth of TPM shells around the hematite particles were monitored at regular intervals using optical microscopy and controlled by introducing additional volumes of TPM ($200\,\mu\mathrm{L}$ at a time)~\cite{van2017preparation}. To expose the hematite cubes, the fully encapsulated particles were dewetted by a chemical trigger~\cite{youssef2016shape} by adding aqueous NaOH to make a total concentration of 0.06 M at 65 $^{circ}$C under gentle stirring. The added base changes the interfacial balance between the cube surface and the TPM oil, so the oil retracts from the hematite and the cube protrudes from the droplet. Dewetting was monitored by optical microscopy until the cubes were visibly extruded, at which point the particles had the desired geometry of a spherical TPM body carrying a single protruding hematite cube (Fig.~\ref{fig1}c). The extent of protrusion was set by the amount of NaOH added and the reaction time.

The dewetted TPM droplets were then hardened by radical polymerisation, which locks in the geometry. A radical initiator, azobisisobutyronitrile (AIBN; $10\,\mathrm{mg}$ per 15 mL), was added to the dispersion, and the sample was kept at $80\,^{\circ}\mathrm{C}$ for $2\,\mathrm{h}$, cross-linking the methacrylate groups of the TPM. After cooling to room temperature, the particles were washed by repeated centrifugation and redispersion in deionised water to remove residual initiator and unreacted species, and were stored in water until use. The composite particles created had the hematite cube as the magnetic ``patch'' and the non-magnetic TPM sphere as a functionalizable, non-interacting part (Fig. \ref{fig1}b and c). The resultant size of the cube-TPM composite was 2.14 $\pm$ 0.04 $\mu$m. The part of the cube protruding out could be either a corner, a face or even an edge. In our experiments, a majority of the particles had a corner protruding from the TPM sphere. We refer to this particle as our `Pacman' particle.

\subsection{Sample preparation and chamber assembly}

The capillary chamber was assembled using two Indium Tin Oxide (ITO) coated glass coverslips (thickness \#1). To minimize non-specific particle-substrate interactions, the coverslips were cleaned through sequential ultrasonic baths: 15 minutes in isopropyl alcohol, 15 minutes in acetone, and a final 15 minutes in isopropyl alcohol. Following sonication, the coverslips were rinsed with deionized water and dried in an oven at 80°C.

The two cleaned ITO coverslips served as the electrodes and were placed in a slightly offset, sandwiched arrangement (Fig. 1a), separated by a 150 $\mu$m parafilm spacer that defined the electrode gap (150 $\mu$m). The staggered positioning provided exposed ITO surfaces for making electrical connections, which were established using silver paste and copper tape (Fig. 1a). 

An aqueous particle suspension was prepared in deionized water at a concentration of 0.01 wt\%. To facilitate smooth introduction into the capillary cell, a low concentration (4.33 mM) of sodium dodecyl sulfate (SDS) surfactant was added.
The colloidal mixture was gently pipetted into the chamber, and the open ends were sealed with a UV-curable adhesive to keep the suspension fully contained.

\subsection{Field Controls and Visualization}
A sinusoidal AC electric field was applied across the cell using a METRAVI DDS-1010 function generator. A magnetic field was provided by a Helmholtz coil assembly, with the sample positioned at its centre to enable in situ magnetic manipulation and real-time observation. Reorienting the Helmholtz coils allowed the magnetic field to be directed either within the x-y plane or along the z-axis. The field strength at the sample's position was verified using a Gaussmeter connected to a Hall probe.

The experimental setup was monitored under an inverted optical microscope using a 100× objective lens in bright-field mode, with dynamic behaviour recorded via a high-speed camera.

\subsection{Simulations}
We performed overdamped Brownian dynamics simulations of $N = 200$ Pacman particles in MATLAB in a $100 \times 100 \times 150 ~ \mu m^3$ box, periodic in $x$ and $y$ with reflective walls in $z$. Each particle carried a point dipole $m = 2.2 \times 10^{-15}$ Am$^2$ along the cube body diagonal.\cite{sacanna2012magnetic, lowrie2007fundamentals} A uniform field along $x$ applied a torque  $m\times B$, and particles interacted via point dipole-dipole forces and torques. Particles also experienced buoyancy-corrected gravity, Brownian translation and rotation in water at 298 K, and a soft repulsion between spheres. Particles were initialized at random non-overlapping positions and orientations, and the system was evolved for 4 hours.

\section{Results and Discussion}

\subsection{Structures formed under purely (in-plane) magnetic fields}

\begin{figure}[h]
	\centering
	\includegraphics[width=8.3 cm]{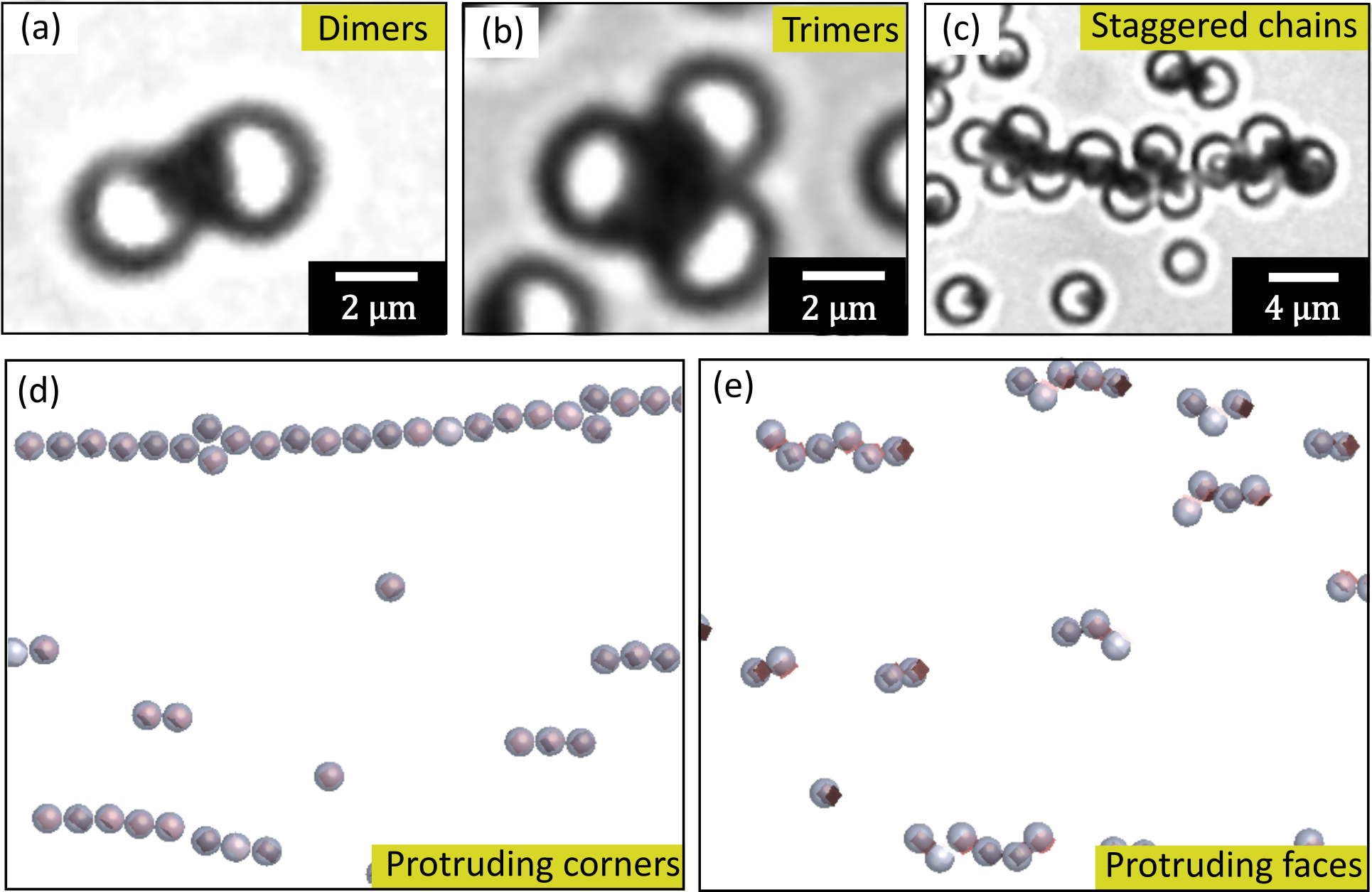}
	\caption{Structures formed due to an applied magnetic field of 1.3 mT along the $x$-axis. (a) Dimers, (b) trimers and (c) staggered chains. The results from Brownian dynamics simulations showing the formation of dimers, trimers and (d) chains when the part of the cube protruding from the tpm sphere is a corner and (e) the formation of staggered chains when the part of the cube protruding out is a face.}
	\label{fig2}
\end{figure}

In water and without external fields, the Pacman particles interact via their magnetic patches, typically forming small clusters of 2-3 particles. When subject to an in-plane magnetic field, the magnetic interactions between the patches are enhanced and, depending on the local density of particles in a particular region, dimers, trimers, staggered chains, and even flocks of particles (that often incorporate smaller staggered chains segments within them) assemble as shown in Fig. \ref{fig2}a-c. The magnetic interactions arise from the canted antiferromagnetic nature of hematite at room temperature (as seen from Fig. \ref{fig1}e, the magnetization value of the hematite cube under zero applied field is 0.3 emu/g). In hematite cubes, the permanent magnetic dipole moment is intrinsically tilted by 12$^\circ{}$ from the [111] body-diagonal direction, introducing a fixed orientational anisotropy within the cubic particle.\cite{rossi2015shape, meijer2013self} Under static magnetic fields, dilute hematite-cube suspensions have been shown to organize into straight or kinked chains depending on field strength \cite{rossi2018self}, while rotating fields drive individual cubes and short chains through rolling, tumbling, and synchronization regimes and, at higher density, produce swarms and a two-dimensional chiral fluid \cite{soni2019odd, petrichenko2020swarming}. 

For the Pacman particle, what plays an essential role is that the magnetic patch is linked with a fairly voluminous TPM sphere, which prevents the cubes from lining up diagonal to diagonal into a long chain, as they normally would, had there not been the TPM sphere attached. In addition, these cubes are oriented randomly with respect to the TPM sphere, but all of them are at the surface, as is seen from Fig. \ref{fig1}b. This is what gives rise to the various particle configurations. Sacanna \textit{et al}. had previously observed the formation of dimers, trimers and colloidal molecules with varying particle sizes.\cite{sacanna2012magnetic} The same off-centre embedding of the hematite core has also been used to engineer size-selective colloidal molecule assembly and magnetically synchronized microgears \cite{abrikosov2013self, aubret2018targeted}. 

In our Brownian Dynamics simulations, we took two cases: one in which all the particles had a random corner of the cube protruding from the TPM sphere and the second in which a random face of the cube protruded from the sphere. In both cases, the extent of the protrusion remained the same. For the protruding corners case, the particles formed longer and well-defined chains (Fig. \ref{fig2}d), whereas for the protruding faces case the chains were shorter and staggered (Fig. \ref{fig2}e). The formation of dimers and trimers is seen in both cases. In real experiments, our system is a combination of these two since we can have either a face or a corner protruding from the sphere.

The dimensionless magnetic dipole-dipole interaction energy is given as:
\begin{equation}
    U^{*}(r_{ij},\mu_{i}^*,\mu_{j}^*)=-(8\mu_{i}^*,\mu_{j}^*)\left[3\frac{(\hat{\mu_{i}}.r_{ij}^*)(\hat{\mu_{j}}.r_{ij}^*)}{r_{ij}^{*5}}-\frac{\hat{\mu_{i}}.\hat\mu_{j}}{r_{ij}^{*3}}\right] 
\end{equation}

where $U^{*}=\frac{U}{k_BT}$, $U$ being the dipole dipole interaction energy, $\mu^*=\chi\mu$ is the dimensionless magnetic dipole moment where $\chi$ is the susceptibility. The scaled distance between particles $i$ and $j$ is $r_{ij}^*=\frac{r_{ij}}{a}$ where $a$ is the particle radius. Abrikosov \textit{et al.}\cite{abrikosov2013self} outlined two important parameters $\beta$ and $s$ in characterizing the position of the cube inside a composite particle. While $s$ denoted the fraction of the radial distance $R$ of the TPM sphere at which the centre of the hematite cube is situated, $\beta$ denoted the angle which the net dipole moment vector of the cube makes with the z axis of the colloidal coordinate frame. The highest probability for structure formation was observed at $\beta=90^\circ$ taking an $s=0.7.$ In our experiments, we see a wide range of values of $s$ as well as different orientations of the cube inside the TPM sphere, indicating a diverse set of $\beta$ values. This explains the appearance of dimers, trimers and staggered chains even at the same particle size.

\subsection{Structures formed under purely (out-of-plane) electric fields}

When subject to an out-of-plane AC electric field according to the geometry shown in Fig. \ref{fig1}f, the structural configurations formed by the colloids depend strongly on the field parameters including frequency and amplitude, and particle parameters like size, composition and zeta potential. If we represent the applied AC electric field with frequency $\omega$ as the real part of the complex phasor $\textbf{E}(t) = E_0 e^{i\omega t}\hat{e_z}$, then the time-averaged (electric) dipole-dipole interaction force on a particle at position \textbf{r} exerted by another particle residing at the coordinate origin is given as: \cite{barros2025structure, gong2017electric}
 \begin{equation}
     \textbf{F}_{E}(\textbf{r})=\frac{3}{4}\pi\epsilon _m a^2K_0^2E_{rms}^2 \left(\frac{2a}{r}\right)^4 \left[(1-3\cos^2{\theta})\hat{r}+\sin{2\theta}\hat{\theta}\right]  
\label{fdip}    
\end{equation}

Here, $K$ denotes the frequency-dependent complex Clausius–Mossotti (CM) factor, expressed as $K=\frac{\tilde{\epsilon}_p-\tilde{\epsilon}_m}{\tilde{\epsilon}_p+2\tilde{\epsilon}_m}$, where $\tilde{\epsilon}_p$ and $\tilde{\epsilon}_m$ are the complex permittivities of the particle and the suspending medium, respectively, and $K_0$ is the magnitude of the CM factor. In this expression, $\epsilon_m$ is the medium permittivity, $a$ is the particle radius, $\textbf{r}$ represents the centre-to-centre separation between particles, $E_{rms}$ is the root-mean-square (rms) value of the applied electric field, and $\theta$ is the angle between the applied field direction and the interparticle axis. For $\theta = 0^\circ$, dipolar interactions lead to attraction between the particles, whereas for $\theta = 90^\circ$, the interaction is repulsive. On the other hand, an electrohydrodynamic (EHD) flow is generated near the surface of a polarizable electrode by an applied electric field. The expression for the EHD force field surrounding a spherical colloidal source particle situated close to an electrode is given as:\cite{ma2015inducing, ristenpart2004assembly}

\begin{equation}
F_{EHD} = \beta \left(\frac{9}{2}\pi \varepsilon_m a\right)\,\kappa h E_{\mathrm{rms}}^2
\frac{
K' + \bar{\omega} K''
}{
1 + \bar{\omega}^2
}
\frac{\left(\frac{r}{a}\right)}{\left[1 + \left(\frac{r}{a}\right)^2\right]^{5/2}}
\end{equation}

where $\bar\omega = \frac{\omega h}{2 \kappa D}$ is the scaled frequency, $\kappa^{-1}$ is the Debye screening length, D is the ion diffusivity, $h$ is the separation between the two electrodes, and $\beta=0.04$ is a scaling factor. $K'$ and $K''$ represent the real and imaginary parts of the Clausius-Mossotti factor, respectively.\cite{ma2015inducing} As discussed in our earlier works,\cite{barros2025structure} the tangential flow velocity, and hence the EHD force, is dependent on the frequency ($\sim \omega ^{-1}$) of the applied E field. At lower frequencies, therefore (5-20 kHz), the EHD force is dominant over the dipolar force. 

The third force that might play a role in the electric field-induced assembly is the dielectrophoretic force (DEP). The time-averaged DEP force on a spherical colloidal particle of radius $a$ is given as: \cite{jones1995electromechanics, morgangreen2003ac}
\begin{equation}     
\langle{\textbf{F}_{DEP}}\rangle = \pi\epsilon_m a^3 K' \nabla\lvert{{E}_{rms}}\rvert^2
\end{equation}
A positive value of $K'$ indicates positive dielectrophoresis; that is, the particles will tend to move towards high-field zones, while a negative value indicates negative dielectrophoresis; that is, the particles will tend to move towards the low-field areas. Therefore, by examining the variation of $K'$ with frequency, the nature of the response to the applied electric field can be understood. Next, we examine the roles of these three forces in forming different structural configurations for a) single-unit particles, that is, either a TPM sphere or a hematite cube and b) composites, that is, the Pacman particles.

\subsubsection{\textbf{Individual assembly of TPM spheres and hematite cubes under electric fields}}

\begin{figure*}[!h]
	\centering
	\includegraphics[width=16 cm]{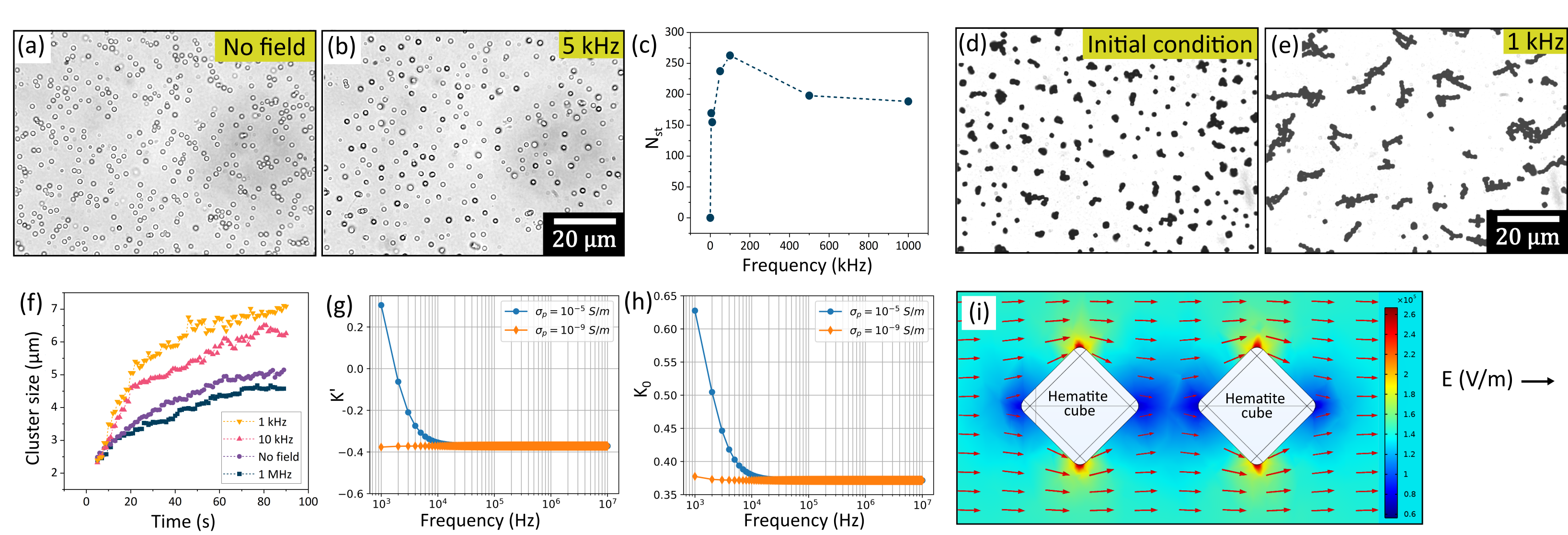}
\caption{Electric-field response of TPM spheres and hematite cubes. (a) TPM spheres in the absence of an applied electric field (b) TPM spheres subjected to an out-of-plane AC electric field at 5 kHz and 20 Vpp, showing the formation of stacks (darker rings) accompanied by particle spreading within the plane. (c) the number of particles participating in stacking (\(N_\mathrm{st}\)) plotted as a function of frequency. (d) Hematite cubes in the absence of an ac electric field and (e) the formation of clusters accompanied by chains formed due to intrinsic magnetic interactions in the presence of an out-of-plane ac electric field of 1 kHz and 20 Vpp. (f) Cluster size as a function of time for hematite cubes subjected to AC electric fields over a range of frequencies. (g) The real part of the Clausius-Mossotti factor for the hematite cubes plotted as a function of frequency for two different conductivity values, i.e. $\sigma = 10^{-5}$ and $10^{-9}$. (h) A similar plot for the absolute value of the Clausius-Mossotti factor and (i) the electric field distribution colour map and quiver plot for two hematite cubes placed edge to edge in an applied AC field of 10 kHz and 20 Vpp from finite element modelling in COMSOL. }
	\label{fig3}
\end{figure*}

In our system, we have a composite particle consisting of two distinct materials (a TPM sphere and a hematite cube) with different dielectric constants immersed in a medium (water) with a third dielectric constant, so estimation of the Clausius-Mossotti factor and consequently prediction of the behaviour of the dipolar, DEP and EHD forces becomes a complex task. Therefore, we first decided to examine the assembly of the individual components, that is, the hematite cubes and TPM spheres separately under an applied AC electric field. TPM spheres alone in deionized water showed the formation of 2-3 particle columnar stacks (seen as circles with darker edges) along with the formation of a non-close-packed, partially ordered lattice upon application of an AC electric field in the z direction (Fig. \ref{fig3}a-b). The Clark-Evans factor calculated for Fig. \ref{fig3}b was $R \approx$ 1.42, which is significantly higher than 1, indicating partial ordering. No significant variation in the assembled structures was observed over a frequency range of 5 kHz-1 MHz except a small variation in $N_{st}$, which is the average number of particles participating in the stacking (Fig. \ref{fig3}c). While the stacking is an outcome of the electric dipolar attraction along the field, the lattice-like configuration perpendicular to the field directly arises from the in-plane inter-particle dipolar repulsion. 

In contrast to the TPM spheres, the hematite cubes alone showed a combined response arising from their own residual magnetization and polarization under an applied electric field. The hematite cubes exhibit canted antiferromagnetism at room temperature and assemble into chains and small clusters due to their residual magnetic moment, even in the absence of any applied field.\cite{morrish1994canted} Therefore, to examine the effect of the electric field, we first applied an out-of-plane magnetic field to break the small clusters via in-plane magnetic dipolar repulsion. If the magnetic field in z is turned off and no electric field is turned on (zero field), the cubes will again start to form small chains and clusters. In our experiment, the perpendicular magnetic field was kept on for 5 seconds (Fig. \ref{fig3}d), following which the magnetic field was switched off and immediately the AC electric field was turned on. This led to the formation of chains and clustered chains (Fig. \ref{fig3}e) whose mean cluster size depended strongly on the applied electric field frequency. As seen in the plot of mean cluster size vs time in Fig. \ref{fig3}f, application of a 1 kHz AC field resulted in the growth of larger chains/clusters in a given time interval, whereas at 10 kHz, the mean cluster size was smaller over the same interval. In the MHz range, the cluster growth became slower than even the zero-field condition, leading to the smallest mean size. This implies that magnetic moment-induced ordering is enhanced in the kHz regime and hindered in the MHz regime as compared to the zero applied field condition. This can be explained by considering that the electrohydrodynamic force (EHD) is expected to persist mainly in the 1 kHz- 15 kHz range, beyond which it falls off as 1/$\omega$. Therefore, in the kHz regime, the EHD-induced clustering cooperates with the magnetic dipolar ordering and leads to the formation of larger clusters in the same time interval. In contrast, in the MHz regime, in-plane electric dipolar repulsion competes with the magnetic attraction and leads to the formation of clusters that are smaller even compared to the zero-field case. 

One important consideration here is the nature of the response of the hematite cubes to the applied electric field. This is explained by first considering the electric field distribution around the hematite cubes and then by considering the frequency dependence of the Clausius-Mossotti (CM) factor. We consider the static dielectric constant ($\epsilon_p$) of bulk hematite to be in the range 15-25,\cite{lunt2013dielectric, young1973compilation} the dielectric constant of water ($\epsilon_m$) to be 80, and the conductivity of water to be $10^{-6}$ S/m. The conductivity of hematite ($\sigma_p$) has a wide range of reported values in the literature, ranging from $10^{-5} - 10^{-9}$ S/m.\cite{morin1951electrical, engel2014electrical, zainuri2017hematite, sahoo2025study} Since the dielectric constant of hematite is much lower than that of water, the hematite particles are basically dielectric `holes' that will repel the electric lines of force. Therefore, as shown in the electric field distribution contour map generated using finite element modelling in COMSOL in Fig. \ref{fig3}i, the field is highest perpendicular to the applied field (red regions), while it is lowest parallel to the field (blue regions). In this plot, the applied field is represented in the horizontal direction. On the other hand, the CM factor for an ellipsoid along the principal axis $n$ is given as:\cite{jones1995electromechanics}

\begin{equation}
    K_{n}(\omega) = \frac{\tilde{\epsilon}_p-\tilde{\epsilon}_m}{3(\tilde{\epsilon}_m+A_n (\tilde{\epsilon}_p-\tilde{\epsilon}_m))}
\end{equation}

where $A_n$ is the depolarization factor. For a cubic particle we can take three equivalent axes such that $A_x=A_y=A_z=1/3$, so that we can write:

\begin{equation}
    K_{cube} = \frac{(\tilde{\epsilon}_p-\tilde{\epsilon}_m)}{ (\tilde{\epsilon}_p+2\tilde{\epsilon}_m)}
\end{equation}

which is similar to that of a spherical particle.  As seen in Fig. \ref{fig3}g, the real part ($K'$) of the CM factor is entirely negative for $\sigma_p = 10^{-9}$ S/m and changes from a positive to a negative value near 2 kHz for $\sigma_p = 10^{-5}$ S/m. This indicates that the hematite cubes are expected to show negative dielectrophoresis ($K'<0$) for the major part of the frequency range. This indicates that the hematite cubes will tend to move into the low-field zones, which in this case is along the direction of application of the field. Consequently, the cubes will tend to line up into columnar stacks parallel to the field purely from dipolar considerations. The fact that we obtain larger in-plane clusters at low frequencies is therefore a direct consequence of the presence of EHD forces in this frequency range. In the MHz regime, while the EHD forces become insignificant, the dipolar forces also weaken. This is observed from the decrease in the absolute value of the CM factor ($K_0$) with frequency leading to a flattening off near 10 kHz for a $\sigma_p$ value of $10^{-5}$ S/m (Fig. \ref{fig3}h). For a $\sigma_p$ value of $10^{-9}$ S/m, a nearly flat response is seen (Fig. \ref{fig3}h). Since the density of hematite is 5 g/cc, the hematite cubes tend to sediment and in the MHz frequency range the dipolar or EHD forces are not strong enough to lift them and assemble them into a column. Consequently, the cubes stay on the x-y plane and experience weakened but repulsive dipolar forces, leading to the formation of smaller clusters.

\subsubsection{Assembly of Pacman particles under electric fields}
\begin{figure}[h]
	\centering
	\includegraphics[width=8.3 cm]{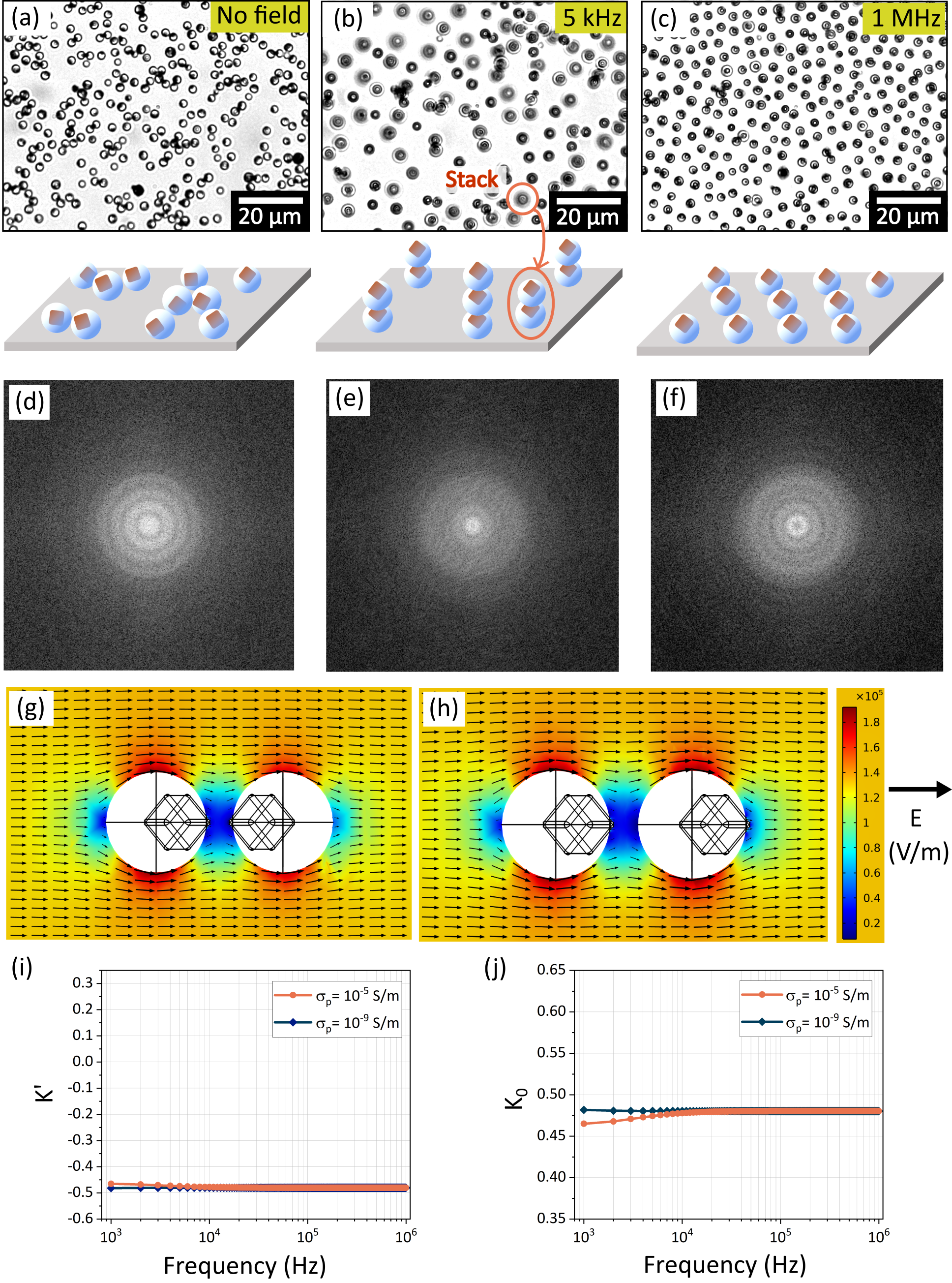}
	\caption{Pacman particles under an applied ac electric field along the $z$-axis. (a) No field condition, (b) at an applied ac frequency of 5 kHz and 20 Vpp (peak-to-peak voltage) showing stacking of the particles in the field direction, (c) at a frequency of 1 MHz and 20 Vpp, the particles separate out in the plane with the cubes facing the viewing objective, forming a non-close-packed lattice-like structure with partial ordering. The schematics are shown below each image. Fast Fourier Transforms (FFTs) of the images of the resultant structures for (d) no field,(e) 5 kHz, and (f) 1 MHz. A colourmap and quiver plot of the electric field distribution for Pacman particles at 5 kHz and 20 Vpp with (g) the cubes facing each other and (h) the cubes facing the same direction, plotted in COMSOL using finite element modelling. (i) The real part and (j) the absolute value of the Clausius-Mossotti factor for the Pacman conjugate for two $\sigma$ values of the hematite cube, i.e. $10^{-5}$ and $10^{-9}$ S/m plotted in te same scale as the CM for hematite in Fig.\ref{fig3}g-h.}
	\label{fig4}
\end{figure}

Compared to TPM spheres or hematite cubes, a drastically different situation arises in the case of Pacman particles. In the absence of any field, the Pacman particles form small clusters because of the attractive magnetic interaction between the hematite patches (Fig. \ref{fig4}a). At low applied electric field frequencies (5-10 kHz) columnar stacks are formed in the direction of the applied field, as shown in Fig. \ref{fig4}b. In bright-field images, the stacks appear as larger, blurry entities due to the presence of a particle(s) above the focal plane. A schematic is shown below Fig. \ref{fig4}a-b for clarity. The columnar stack formation process is also visible from the supplementary video V2. However, at high frequencies (~1 MHz), the columnar stacks disintegrate and the Pacman particles spread out into non-close-packed, mutually repelling, lattice-like configurations as shown in Fig. \ref{fig4}c and the schematic below, with a Clark-Evans factor $R \approx$ 1.6. In the inverted microscope image, the hematite cubes appear to have their faces oriented upward, indicating that they are in contact with the bottom electrode surface. This orientation is likely favoured by gravity, as placing the denser hematite cube closer to the substrate lowers the height of the centre-of-mass of the Pacman particle. The appearance of the partially ordered lattice at 1 MHz is further demonstrated by the FFT images (Fig. \ref{fig4}d-f) corresponding to the bright field images above. While at 1 MHz distinct, well-separated rings are observed, at 5 kHz, a diffused pattern is seen. 

The structural configurations formed by the Pacman particles under an applied electric field can be understood in the light of the behaviours of the individual TPM and hematite colloids. The electric field distribution around the Pacman particles was simulated by us using finite element modelling in COMSOL (Fig. \ref{fig4}g-h), assuming a) one corner of the cube is protruding from the sphere in the majority of the cases and b) the centre of the cube sits at a position which is 50\% of the radius of the TPM sphere. The edges of the cubes have been rounded to match the real particles. The high field zones were observed perpendicular to the applied field and low field zones appeared parallel to the field as before, but these zones extended over larger areas as compared to the hematite cubes. The real and absolute values of the Clausius--Mossotti (K) factor calculated using COMSOL for the same two hematite particle conductivities, $10^{-5}$ and $10^{-9}$ S/m, are shown in Fig. \ref{fig4}i-j, respectively, plotted on the same scale as the corresponding CM factor plots for hematite cubes in Fig. \ref{fig3}g-h. At this scale, both quantities appear nearly flat over the frequency range considered. The real part of the CM factor remains negative, indicating negative dielectrophoresis, while its absolute value exhibits little variation with frequency. We hypothesize that, within the composite particle, the TPM component drives both columnar stacking and the formation of in-plane repulsive lattices, while the hematite component exhibits a distinct frequency-dependent interplay of interactions: magnetic and EHD interactions cooperate in the low-frequency regime (5-10 kHz), whereas magnetic and electric dipolar interactions compete at high frequencies (1~ MHz). Therefore, in the kHz range, the attractive interaction among the individual Pacman particles (through their magnetic patches) is greater, leading to a large number of them getting in close proximity to each other and consequently favouring the formation of columnar stacks in the direction of the applied field due to the attractive electric dipolar interactions. On the other hand, in the MHz regime, the magnetic attraction between the hematite patches is effectively hindered due to the electric dipolar repulsion in-plane, due to which the Pacman particles prefer to form a partially ordered, non-close-packed lattice-like configuration on the viewing plane. Our experimental observations indicate that the effects of the hematite cube and TPM sphere in a composite Pacman particle are effectively additive. The main role of the structural anisotropy of the Pacman particle lies in the modification of the electric field magnitude distribution, while still being a negative dielectrophoretic system.

\subsection{Reconfigurable Structures from Combined Electric and Magnetic Fields}

\begin{figure}[h]
	\centering
	\includegraphics[width=8.3 cm]{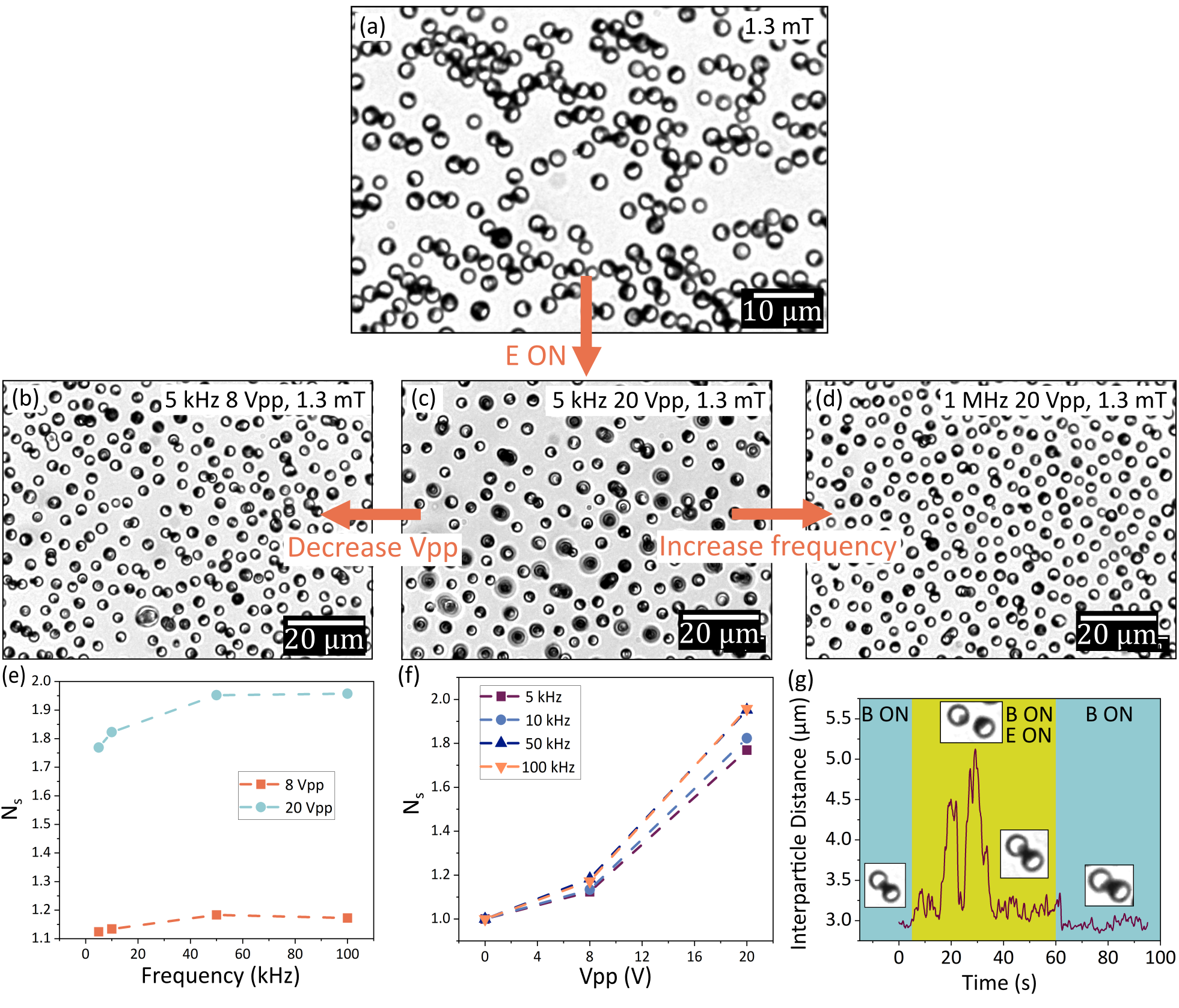}
	\caption{Combined fields: (a) Linear chains and small clusters formed under a magnetic field of 1.3 mT alone. (b) Structures obtained under a combined magnetic field of 1.3 mT and an AC electric field of 8 Vpp at 5 kHz. (c) The stacked structures in (b) increase in size when the electric field strength is increased to 20 Vpp. (d) The planar, non-close-packed ordering observed at 1 MHz and 20 Vpp. The number of particles per stack plotted as a function of (e) frequency and (f) applied voltage (Vpp) shows an increasing trend with Vpp and frequency, saturating beyond 50 kHz. (g) Interparticle distance as a function of time for a spontaneously reconfiguring dimer, showing repeated attachment and detachment events.}
	\label{fig5}
\end{figure}

Building on our understanding of the individual responses of the Pacman particles to electric and magnetic fields, we subsequently investigated field-directed reconfiguration by combining electric and magnetic fields to dynamically control the assembled structures. We used a combination of an electric field (applied along the $z$ direction) and a magnetic field ($B = 1.3~\mathrm{mT}$ applied along the $x$ direction). As before, under only the $B$ field, chains and clusters were observed (Fig.~\ref{fig5}a). Under an applied AC electric field of $E = 20~\mathrm{Vpp}$ with a frequency of 5~kHz, smaller and sparser stacks (Fig.~\ref{fig5}c) were formed in comparison to the purely electric case (Fig.~\ref{fig4}b). At lower voltages (8~Vpp), the stack size as well as number of stacks was even smaller (Fig.~\ref{fig5}b). At higher frequencies (1~MHz), the particles still spread out, though less uniformly than before, and often retained dimers due to residual magnetic attraction (Fig. \ref{fig5}d). The competition between electric and magnetic interactions provides a powerful means of controlling the assembled structures. By tuning the relative strengths of the electric and magnetic fields, together with the frequency of the applied electric field, we can not only switch between a stacked columnar state and a dispersed repulsive lattice, but also continuously regulate the extent and degree of structural organization. To illustrate this, we calculated the average number of particles in each stack ($N_{s}$) as a function of frequency and voltage. As shown in Fig. \ref{fig5}e, $N_{s}$ increases with increasing frequency and subsequently saturates at frequencies up to 100 kHz, indicating the progressive formation of larger columnar stacks. At higher frequencies, beyond 500 kHz, the columnar organization is suppressed, and the formation of a purely in-plane repulsive lattice becomes increasingly favourable. $N_{s}$, as expected, is higher for 20 Vpp as compared to 8 Vpp for the same applied frequencies. $N_{s}$ also increases progressively with the applied voltage (Fig. \ref{fig5}f), the effect being more pronounced as the frequency is increased up to 100 kHz. The fine-tuning of the field parameters therefore gives us precise control over the structure formation.

In a multi-field system, there is a competition between $E$- and $B$-field-based ordering, which implies that the $B$-field-induced linear chaining prevents the formation of $E$-field-induced columnar stacking/repulsive lattice formation. Sometimes this leads to a delicate balance, where a spontaneously reconfiguring structure is observed. Under a constant $B$ field of 1.3~mT and an $E$ field of 10~kHz and 8~Vpp, we observed that dimers formed via magnetic interactions repeatedly broke and reassembled due to a balance between magnetic attraction and electric-field-based repulsion (supplementary video V3). By tracking the particle trajectories and quantifying the inter-particle distance over time, we demonstrated this continuously occurring spontaneous reconfiguration (Fig.~\ref{fig5}g). The peaks in the green region (both E and B field ON) in Fig.~\ref{fig5}g show the repeated formation and breaking of the dimer, while the blue region shows the stable dimer configuration without the $E$ field (B ON).

\begin{figure}[h]
	\centering
	\includegraphics[width=8.3 cm]{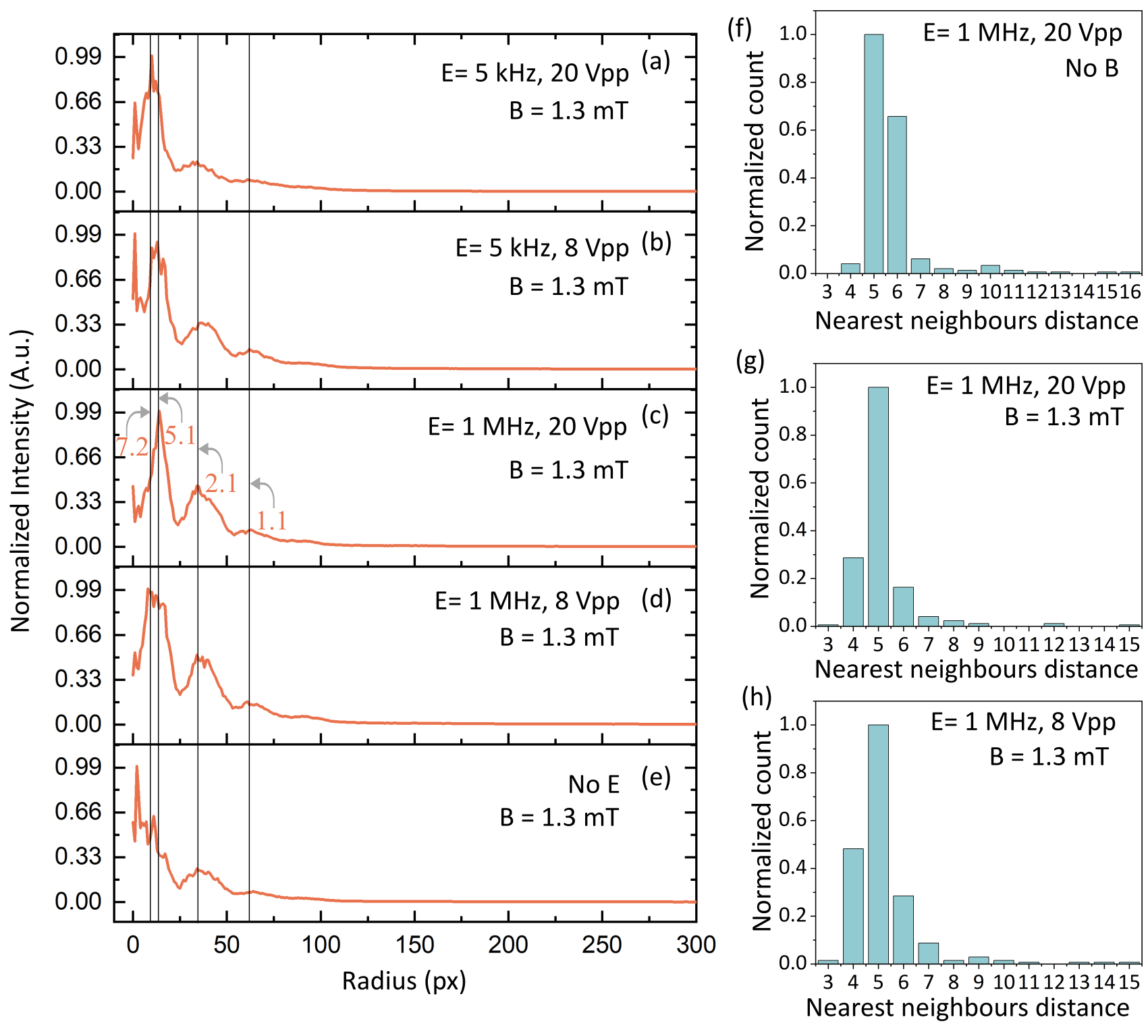}
	\caption{The radial profile plots of the FFT for Pacman particles under different field conditions: (a) an out-of-plane AC electric field of 5 kHz and 20 Vpp combined with an in-plane magnetic field of 1.3 mT; (b) the same field conditions with the electric-field amplitude reduced to 8 Vpp; (c) an out-of-plane AC electric field of 1 MHz and 20 Vpp combined with an in-plane magnetic field of 1.3 mT; (d) the same conditions with the electric-field amplitude reduced to 8 Vpp; and (e) an in-plane magnetic field of 1.3 mT in the absence of an electric field. Histograms of nearest-neighbour distances obtained from Voronoi analysis for (f) an out-of-plane AC electric field of 1 MHz and 20 Vpp without a magnetic field, (g) the same electric-field conditions combined with an in-plane magnetic field of 1.3 mT, and (h) a reduced electric-field amplitude of 8 Vpp with the same magnetic field. The nearest-neighbour distance distribution is narrowest in the absence of the magnetic field and broadens upon its application, with a further increase in spread at reduced electric-field amplitude. This broadening indicates a progressive loss of structural order as the relative influence of the magnetic field increases. }
	\label{fig6}
\end{figure}

To quantitatively illustrate reconfigurability under competition of electric and magnetic forces, we obtained the Fast Fourier Transform (FFT) of each image of the final structures formed under a range of applied voltages and frequencies while the magnetic field was kept constant at 1.3 mT. Subsequently, the normalized radial intensity profile of the FFT was plotted as shown in Fig. \ref{fig6}a-e. From the peak positions (in pixels) of the four observed peaks, the corresponding length scales were calculated by the formula,
\begin{equation}
\text{Length scale} = \frac{N \times 0.12}{\text{position}},
\end{equation}
where $0.12 \,\mu\text{m/pixel}$ is the image calibration factor and $N \times N$ is the size of the FFT image in pixels. The length scales corresponding to these peaks were $L =$7.2 $\mu$m, 5.1 $\mu$m, 2.1 $\mu$m and 1.1 $\mu$m, which are marked by dark lines in Fig. \ref{fig6}a-e. The $L=$ 1.1 $\mu$m peak corresponds to the size of the cube (1 $\mu$m) that is encapsulated in the TPM. This peak appears fairly consistently in all the cases except in (a), where enhanced stacking led to poor visibility of the cubes. The $L=$ 2.1 $\mu$m peak corresponds to the size of the conjugate (Pacman) particle. The $L=$ 5.1 $\mu$m peak is strongest for case (c), which has the highest frequency and voltage, and corresponds to the average separation between the particles in a repulsive lattice. Reducing the electric field amplitude to 8 Vpp (case (d)) broadens this peak while also splitting it into three small peaks corresponding to length scales between 4-7 $\mu$m, indicating a spread in the distribution of distances between particles and the formation of dimers (4 $\mu$m). 

To further quantify the particle spacing, we used a Voronoi analysis, which determines nearest neighbours by considering each particle at the centre of one Voronoi cell, and eliminates any subjectivity in selection of nearest neighbours. We calculated the average nearest neighbour distances for each particle and plotted the normalized histograms (Fig. \ref{fig6}f-h) for three cases: (f) E = 1 MHz, 20 Vpp; B = 0, (g) E = 1 MHz, 20 Vpp; B = 1.3 mT and (h) E = 1 MHz, 8 Vpp; B = 1.3 mT. It is clearly seen that the nearest neighbour distance of 5 $\mu$m is most prominent for the case (f) with maximum E-field amplitude and no B field at 1 MHz (Fig. \ref{fig4}c ). Under a B field, a larger spread around 5 $\mu$m is observed, as seen from Fig. \ref{fig6}g-h and the standard deviation plot in Supplementary Fig. S1. The spread is maximum when the amplitude of the E field is reduced to 8 Vpp (Fig. \ref{fig6} h) and B field is at 1.3 mT. The Voronoi analysis together with the FFT-based radial intensity profile succinctly illustrates the fine-tuning of the assembly under combined fields. This interplay between the electric and magnetic forces therefore allows us to create a variety of reconfigurable structures out of anisotropic building blocks, offering promising opportunities for developing stimuli-responsive smart materials.

\section*{Conclusion}
The assembly behaviour of Pacman particles- hematite and TPM conjugates was studied under crossed electric and magnetic fields. By tuning the delicate balance between the electric and magnetic fields, and harnessing the selectivity of the magnetic patch, a variety of structures were obtained, such as chains, stacks, dimers, trimers, non-close-packed semi ordered lattices, flocks and even a combination of these structures, such as chains in flocks, stacks in lattices and even dimers in stacks or lattices. This paves the way to a rich composition of structures for the design of reconfigurable or switchable smart materials controlled externally by electric or magnetic fields.

\section*{Author contributions}
IB and IS fabricated the sample chambers and did the field-based experiments. IB performed the data analysis, COMSOL modelling and Brownian Dynamics simulations. IB, and IC contributed to the planning of the experiments and writing of the paper.

\section*{Conflicts of interest}
There are no conflicts to declare.

\section*{Data availability}
Most of the data used in this work are either in the main paper. Additional data will be available from the authors upon request, as most of these are video data and are too bulky to store in a repository.

\section*{Acknowledgements}
We acknowledge Prof. Toby Joseph of BITS- Goa for discussions and guidance on the Brownian Dynamics Simulations. We acknowledge the Start-up grant (SRG/2021/001696) of the Science and Engineering Research Board (SERB) (at present the Anusandhan National Research Foundation (ANRF)), Government of India and the additional competitive research grant (GOA/ACG/2021-2022/Nov/03) from BITS-Pilani for this work. We also acknowledge the Central Sophisticated Instrumentation Facility (CSIF) at BITS-Goa for the XRD measurements and the Department of Physics at BITS-Goa for the VSM measurements.



\balance


\bibliography{rsc} 
\bibliographystyle{rsc} 

\end{document}